\documentclass[a4paper,11pt]{article}
\usepackage[parfill]{parskip}
\usepackage{physics, tensor, float, subcaption}
\usepackage{lineno}
\usepackage{jheppub}
\usepackage{amssymb,amsmath,amsthm}
\usepackage{mathrsfs}
\usepackage[utf8]{inputenc}
\usepackage{enumerate}
\usepackage{bigints}
\usepackage{float}
\usepackage{tikz}
\usepackage{setspace}
\usepackage{cancel}
\usepackage{array}
\usepackage{xcolor}
\usepackage{tabulary}
\usepackage{doi}
\usepackage{comment}
\allowdisplaybreaks

\definecolor{lime}{HTML}{A6CE39}
\newcommand{\orcidicon}{%
	\begin{tikzpicture}
	\draw[lime, fill=lime] (0,0) 
		circle [radius=0.16] 
		node[white] {{\fontfamily{qag}\selectfont \tiny ID}};
	\draw[white, fill=white] (-0.0625,0.095) 
		circle [radius=0.007];
	\end{tikzpicture}
	\hspace{-5mm}
}
\newcommand\orcidRudeep{{\href{https://orcid.org/0009-0002-0162-562X}{\orcidicon}}}

\newcommand\orcidChris{{\href{https://orcid.org/0009-0002-6751-2695}{\orcidicon}}}

\newcommand\orcidMarco{{\href{https://orcid.org/0000-0003-2783-3603}{\orcidicon}}}

\newcommand\numberthis{\addtocounter{equation}{1}\tag{\theequation}}

\definecolor{DRed}{rgb}{0.71,0.14,0.07}

\newcommand{\be}{\begin{equation}}
\newcommand{\ee}{\end{equation}}

\def\d{{\mathrm{d}}}
\newcommand{\B}[1]{{\Bar{#1}}}
\newcommand{\T}[1]{{\check{#1}}}

\newcommand{\laund}[1]{\mathcal{O}{\left(#1\right)}}

\def\d{{\mathrm{d}}}
\protected\def\verythinspace{%
  \ifmmode
    \mskip0.5\thinmuskip
  \else
    \ifhmode
      \kern0.08334em
    \fi
  \fi
}

\newcommand{\thin}{\verythinspace}

\begin{document}

\title{\vspace{-25pt}\huge{Kerr--Newman Memory Effect}}

\author[a]{Marco Galoppo\!\orcidMarco,}
\author[b]{Rudeep Gaur\!\orcidRudeep,}
\author[a]{Christopher Harvey-Hawes\!\orcidChris}

\affiliation[a]{School of Physical \& Chemical Sciences, University of Canterbury,
\null Private Bag 4800, Christchurch 8041, New Zealand.}
\affiliation[b]{School of Mathematics and Statistics, Victoria University of Wellington, 
\null PO Box 600, Wellington 6140, New Zealand.}

\emailAdd{marco.galoppo@pg.canterbury.ac.nz}
\emailAdd{rudeep.gaur@sms.vuw.ac.nz}
\emailAdd{christopher.harvey-hawes@pg.canterbury.ac.nz}
\abstract{
\vspace{1em}

We bring the Kerr--Newman spacetime into the Bondi--Sachs gauge by means of zero angular momentum, null geodesics. We compute the memory effect produced at the black hole horizon by a transient gravitational shock wave, which from future null infinity is seen as a Bondi-Metzner-Sachs supertranslation. This results in a change of the supertransformation charges at infinity between the spacetime geometries defined by the black hole before, and after, the shockwave scattering. For an extremal Kerr--Newman black hole, we give the complementary description of this process in the near-horizon limit, as seen by an observer hovering over the horizon. In this limit, we compute the supertranformation charges and compare them to those calculated at null infinity. We analyze the effect of these transformations on the electromagnetic gauge field and explore the self-interaction between this and the angular momentum of the black hole. 
}

\maketitle

\section{Introduction}

The Kerr--Newman spacetime \cite{newmanMetricRotatingCharged1965} describes a rotating, charged Black Hole (BH) and represents the most general of the asymptotically Minkowskian, stationary BH solutions to the Einstein--Maxwell equations. It is a direct generalisation of the Kerr solution \cite{kerrGravitationalFieldSpinning1963} for a chargeless, rotating BH. The Kerr solution is widely accepted as providing an accurate description of the exterior spacetime surrounding realistic BHs. In particular, the matching of its ray tracing predictions with the recent direct observations of Sagittarius A* and M87* further support its relevance \cite{EventHorizonTelescopeCollaboration_2022,EHT_tests, EHT_M87_I,EHT_M87_II}. In spite of these successes, a more realistic representation of a BH would have to include the effects of its inherent electromagnetic charge, in principle varying over time due to the in-fall of charged matter. The Kerr--Newman solution represents a first step in this direction, as it provides the spacetime geometry for an intrinsically charged, stationary, rotating BH. As such, the study of this solution is of critical interest for understanding the dynamics and structure of physically realistic BHs. 

In particular, studying the asymptotic structure of Kerr--Newman BHs may provide further insight into the gravitational memory effect -- i.e., the permanent alteration of the system due to the passage of a gravitational wave \cite{gravmem1st,chriswork1,Thorne:1992sdb,metricofsimpleexample,nonlineargraviationalwave,Tolish,tolishthesis,Garfinkle:2022dnm} -- and its relation to the set of asymptotic infinite symmetries at null infinity, dubbed supertransformations \cite{strominger2016gravitational,compere_lectures,stromingernotes,HPS,HPSrot}. This group of symmetries, composed of \emph{supertranslations} and \emph{superrotations}, was introduced by Bondi, van der Burg, Metzner and Sachs \cite{bondi1960gravitational,Bondi:1962px,Sachs:1962wk,cit:sachs1}, thus it is known as the BMS group. A complete understanding of the BMS group in physically relevant spacetimes is of great interest, as the initial and final states of two particles left permanently displaced by a gravitational wave are related by supertranslations at null infinity. Specifically, in light of the next generation of gravitational wave detectors soon to be operational -- e.g., the Einstein Telescope \cite{ET_Branchesi:2023mws} and the Laser Interferometer Space Antenna (LISA) \cite{Lisa,LISA:2022kgy} -- the memory effect may soon be observationally detected \cite{ET_Goncharov:2023woe}. 
Furthermore, recent development in the study of the BMS algebra and its associated charges \cite{Barnich:2009se,Barnich_troes,aspectsofbms/cft} have led to interesting insight into the ``scattering" problem in general relativity \cite{strominger2016gravitational,stromingernotes} and in the information loss problem \cite{Strominger:2017aeh,HPSrot,HPS}.\\

Moreover, a regime of interest is the Near Horizon (NH) limit of the Kerr--Newman spacetime, specifically the case of an extremal Kerr--Newman BH. Indeed, the study of the NH limit of classical BHs is of fundamental importance for the investigation of their geometry and topology, carrying consequences for any traditional approach to quantum gravity \cite{compere_kerrcft_2012,kunduri_classification_2013-2}. Here, extremal BHs are highly relevant because even as semi-classical objects, they remain inert, since they do not emit any Hawking radiation \cite{bardeen_four_1973,hawking_particle_1975-1}. As such, they represent simple objects for investigating links between quantum physics and general relativity. Furthermore, the NH limit gives a framework for describing the gravitational shockwave scattering as seen by an observer hovering over the horizon. Therefore, it provides a complementary analysis to the memory effect study carried out at null infinity. For such cases as the Reisnner Nordstr\"om and Kaluza--Klein BHs, the NH observer is known to measure a horizon superrotation after the scattering process has occurred --- something absent at null infinity \cite{black_hole_memory_og,KKSofthair}. Moreover, the passage of a gravitational shockwave imparts \emph{soft electric hairs} on the horizon of a charged BH, thus showing the interplay between the gravitational and electromagnetic fields. We will reproduce these calculations for the near horizon extremal Kerr--Newman BH, and further show that the interaction between angular momentum and the electromagnetic field is present even for the bald extremal Kerr--Newman solution. 

The structure of this paper is as follows. We will begin by summarising the expanded Bondi–Sachs (BS) metric, asymptotic Killing vectors, their respective symmetries and associated charges, and the relationship between the memory effect and supertranslations. In Section \ref{sec:2} we put the Kerr--Newman metric in the Bondi-Sachs gauge. In Section \ref{sec:3} we supertranslate the resulting spacetime, electromagnetic gauge field and discuss the physical implications of this procedure in the presence of charge. In Section \ref{sec:4} we explore NH physics for an extremal Kerr--Newman BH and relate the effect of outgoing gravitational radiation to null infinity with the respective modifications of the BH horizon. Section \ref{sec:5} presents a brief summary of the results and a discussion regarding future lines of research. 


\subsection{The Bondi--Sachs Metric}

A natural framework for the investigation of the memory effect is given via casting the spacetime into the Generalised Bondi--Sachs (GBS) form. This metric form was introduced by Bondi, van der Burg, Metzner, and Sachs \cite{bondi1960gravitational,Bondi:1962px,Sachs:1962wk,cit:sachs1} in an attempt to define a concept of asymptotic flatness at null infinity. The falloffs needed to be restrictive enough that unphyiscal spacetimes --- such as those with infinite energy --- would be ruled out, yet not so restrictive that physical spacetimes and gravitational waves would be excluded. In the canonical BMS framework, employed throughout the paper, the falloffs are  \cite{stromingernotes}
\begin{align}\label{falloffs}
    &g_{uu}=-1 +\mathcal{O}(r^{-1}),\quad g_{ur} = -1 + \mathcal{O}(r^{-2}),\quad g_{uA}= \mathcal{O}(r^{0}),\quad g_{AB}= r^2\gamma_{AB} + \mathcal{O}(r)\, ,
\end{align}
with the additional constraints
\begin{equation}\label{eq:constraints}
    g_{rr}=g_{rA}=0 \,.
\end{equation}
The class of allowed asymptotic line elements for these falloffs can be cast in the form
\begin{align}\label{generalBMsexpansion}
  \dd s^2 = &-\dd u^2 -2\thin \dd u\thin \dd r + r^2\gamma_{AB} \thin \dd \Theta^A \dd \Theta^B \nonumber\\ 
  &+\frac{2\thin m_{\,\text{bondi}}}{r} \dd u^2 +r \thin C_{AB}\thin \dd \Theta^A\dd\Theta^B + D^BC_{AB}\thin\dd u\thin \dd\Theta^A \nonumber\\
  &+\frac{1}{16\thin r^2}\thin \Big\{C^{FD}C_{FD}\Big\}\,\dd u\thin \dd r \nonumber\\
  &+\frac{1}{r}\Big(\,\frac{4}{3}N_A +\frac{4u}{3}\partial_Am_{\,\text{bondi}}-\frac{1}{8}\partial_A\Big\{C^{FD}C_{FD}\Big\}\Big)\dd u\thin \dd\Theta^A \nonumber\\
  &+\frac{1}{4}\gamma_{AB}\thin \Big\{C^{FD}C_{FD}\Big\}\,\dd \Theta^A \dd \Theta^B + \dots
\end{align}
where $\Theta^A \in \{\theta,\phi\}$ and the uppercase Latin indices run over $\theta,\phi$. $D_A$ is the covariant derivative on the 2--sphere with respect to the 2--sphere metric, $\gamma_{AB}$. The function $m_{\text{bondi}}$ is the Bondi mass aspect, which is in general a function of $u,\, \theta$, and $\phi$. Its integral over the 2--sphere at null infinity gives the total Bondi mass for the spacetime. In the case of the Kerr--Newmann spacetime, the Bondi mass coincides with the canonical mass of the black hole. $N_A$ is the angular moment aspect. Contracting $N_A$ with the generator of rotations and integrating over the entire sphere then gives the total angular momentum of the spacetime. $C_{AB}$ is another field, symmetric and traceless ($\gamma^{AB}C_{AB} = 0$). The retarded time derivative of $C_{AB}$ defines the Bondi news tensor, 
\begin{equation}
    N_{AB} := \partial_{u}\thin C_{AB}\,.
\end{equation}
The news tensor is the gravitational analogue of the Maxwell field strength and its square is proportional to the gravitational energy flux across $\mathcal{I}^+$ (future null infinity) \cite{stromingernotes,compere_lectures}. We note that $N_{A}$ is defined slightly differently in various parts of the literature --- usually depending on the asymptotic expansion \eqref{generalBMsexpansion}. We conform to the convention employed by Strominger \cite{stromingernotes} and Comper\'e \cite{compere_lectures}, which uses the decomposition \eqref{generalBMsexpansion}, leading to $N_A$ being defined as
\begin{equation}\label{stromingerNAdef}
    \frac{2}{3}N_A -\frac{1}{16}\partial_{A}\Big(C_{BC}C^{BC}\Big):= g^{(1)}{_{uA}}\,. 
\end{equation}
Here, $g^{(1)}_{uA}$ corresponds to the $r^{-1}$ expansion in $g_{uA}$. 


\subsection{Asymptotic Killing Vectors}

The symmetries of a spacetime are directly associated to its Killing vectors. Thus, before discussing the charges associated with the BMS asymptotic symmetries, we must introduce its asymptotic Killing vectors. The most general Killing vector, $\xi^{\alpha}$, that preserves the metric \eqref{generalBMsexpansion} to leading order is \cite{stromingernotes}
\begin{align}\label{Full_Asymptotic_Killing_Vec}
    \xi^{\alpha}\partial_{\alpha} := f\partial_u &+ \Bigg[-\frac{1}{r}D^Af + \frac{1}{2\thin r^2}C^{AB}D_{B}f + \laund{r^{-3}}\Bigg]\,\partial_A\\
    &+\Bigg[\frac{1}{2}D^2f -\frac{1}{r}\Bigg\{\frac{1}{2}D_Af D_B\thin C^{AB} + \frac{1}{4}C^{AB}\thin D_{A}D_B \thin f\Bigg\}+ \laund{r^{-2}} \Bigg]\,\partial_r\,,
\end{align}
where $f$ is a function of that angular coordinates $(\theta,\phi)$ only and $D^2$ is the standard Laplacian on the 2--sphere. However, since we conduct an analysis to the leading order --- as done in \cite{HPSrot}, the Killing vector is truncated
\begin{equation}\label{Asymptotic_Killing_vec_truncated}
    \xi^{\alpha}\partial_{\alpha} = f\thin \partial_u + \frac{1}{2}D^2 f \partial_r - \frac{1}{r}D^Af\partial_A\,.
\end{equation}


\subsection{Associated Charges and Charge Conservation}

It is well known that the symmetries of spacetime are associated to conserved charges via Noether's theorem. It follows that the BMS group, which is an infinite dimensional group of symmetries at null infinity, also has an infinite number of charges associated to supertranslations and superrotations. The supertranslation charge and its conservation is given by \cite{stromingernotes}
\begin{equation}\label{supertranslation_charge}
    Q_f^+ = \frac{1}{4\pi} \int_{\mathcal{I}^+} \dd^2 \Theta \,\sqrt{\gamma}\,fm_{\,\text{bondi}} =  \frac{1}{4\pi} \int_{\mathcal{I}^-} \dd^2 \Theta\, \sqrt{\gamma}\,fm_{\,\text{bondi}} = Q^-_f\,,
\end{equation}
where the integration is carried out over the 2--sphere at null infinity, and $f$ is a function of angular coordinates that can be thought of as the generator of supertranslations. In general, the supertranslation charges will depend on advanced/retarded time, following the respective dependence of $m_{\text{bondi}}$. For $f=1$ the conserved charge is the spacetime energy whilst in the case where $f$ is a $l=1$ spherical harmonic function we have conservation of ADM momentum.

The superrotation charge and its conservation is given by \cite{stromingernotes}
\begin{equation}\label{superrotation_charge}
    Q^+_Y = \frac{1}{8\pi}\int_{\mathcal{I}^+} \dd^2\Theta \,\sqrt{\gamma}\, Y^A N_A = \frac{1}{8\pi}\int_{\mathcal{I}^-} \dd^2\Theta\, \sqrt{\gamma}\, Y^A N_A = Q^-_Y,
\end{equation}
where $Y^{A}$ is an arbitrary vector field on the 2--sphere. If $Y^A$ is chosen to be one of the 6 global conformal Killing vectors on the 2--sphere, then \eqref{superrotation_charge} expresses the conservation of ADM angular momentum and boost charges. Moreover, the notation employed indicates that the supertransformation charge at future null infinity, $\mathcal{I}^+$, will match the supertransformation charge at past null infinity, $\mathcal{I}^-$. These matching conditions are a requirement for a well-posed scattering problem of gravitational waves in general relativity.


\subsection{Memory Effect and Supertranslations}

The direct correspondence between the gravitational memory effect and the action of supertransformations of the BMS group has been a subject of ongoing debate \cite{stromingernotes,HPSrot,HPS,compere_lectures,Compere_final_state,donnaysuperathorizon,extendedsymm,Wald_bHM}. For example, in the case of a Schwarzchild BH there is an approximate mapping between the action of a BMS supertranslation at null infinity and a gravitational shockwave \cite{HPSrot}. This mapping has not been shown for more general cases e.g. for the Kerr and Kerr--Newman BH solutions. However, this identification is believed to be consistent throughout the classical family of BH solutions. As such, we assume that this physical mapping proven for non-rotating BHs can also be applied to rotating cases. A supertranslation is given by taking the Lie derivative of the spacetime metric along the asymptotic Killing vector
\begin{equation}\label{asymptoticKillingVector}
    \xi^{\mu} \partial_{\mu} = f\partial_{v} - \frac{1}{2}D^2 f\partial_r + \frac{1}{r}D^{A}f\partial_A \,.
\end{equation}
As shown in \cite{HPSrot}, the action of such a supertranslation on the Schwarzchild metric can only equate to part of the deformation a gravitational wave would produce when striking the BH. Indeed, the full memory effect due to the gravitational shockwave is written in this case as
\begin{equation}\label{eq:real_change}
    \delta g_{\mu\nu} = \mathcal{L}_{\xi} g_{\mu\nu} + \frac{2\mu}{r}\delta^{\thin v}{_{\mu}}\delta^{\thin v}{_{\nu}}\,,
\end{equation}
where $\delta g_{\mu\nu}$ refers to the permanent change in the spacetime due to the passage of a gravitational wave. Notice that the second term on the r.h.s. of \eqref{eq:real_change} is to be expected, as gravitational waves carry energy, linear and angular momentum. Therefore, the shockwave scattering on the BH will bring about changes in these quantities in both. Hence, focusing on the case discussed in \cite{HPSrot}, the mass of the hairy Schwarzschild BH would be $m = M + \mu$ after the gravitational shockwave strikes the BH. However, the Bondi mass of the BH --- at linear order --- does not change. Thus emphasising that, in general, the memory effect is not entirely captured by BMS supertransformations, although it is seen as so if studied from null infinity.

 
\section{Kerr--Newman Spacetime in the Bondi--Sachs Gauge} \label{sec:2}

The Kerr--Newman line element in Boyer--Lindquist coordinates $\qty{\B{t},\B{r}, \B{\theta}, \B{\phi}}$ is
\begin{equation}\label{eq:Kerr--Newman_Boyer--Lindquist}
    \dd s^2 = - \qty( \frac{\dd{\Bar{r}^2}}{\Bar{\Delta}} + \dd{\Bar{\theta}^2} ) \thin \Bar{\rho}^2 + ( \dd{\Bar{t}} - a \sin^2\Bar{\theta} \dd{\Bar{\phi}} )^2 \thin \frac{\Bar{\Delta}}{\Bar{\rho}^2} - (\Bar{A}^2 \dd{\Bar{\phi}} - a  \dd{\Bar{t}})^2 \thin \frac{\sin^2 \Bar{\theta}}{\Bar{\rho}^2},
\end{equation}
with
\begin{align}
   & \Bar{\Delta}(\B{r}) = \B{r}^2 + a^2 - \Xi, \\
   & \B{\Xi}(\B{r}) = 2 \thin M \, \Bar{r} - Q^2, \label{Qpirck}\\ 
   & \Bar{\rho}^2(\B{r},\B{\theta}) = \Bar{r}^2 + a^2 \cos^2 \Bar{\theta}, \\
   & \Bar{A}^2(\B{r}) = \Bar{r^2} + a^2 ,
\end{align}
where $M,a$ and $Q$ are the mass, angular momentum per unit mass, and electric charge of the Kerr--Newman black hole in geometrised units, respectively. We aim to cast \eqref{eq:Kerr--Newman_Boyer--Lindquist} into the BS gauge. To do so, we have to first move from the Boyer--Lindquist coordinates to the GBS coordinates, in which the metric has to respect the constraints \eqref{eq:constraints}, and then impose the falloffs \eqref{falloffs} through a further coordinate transformation. Only once \eqref{eq:Kerr--Newman_Boyer--Lindquist} is put into the BS gauge, then a meaningful analysis of the asymptotic structure can be made possible. 

In analogy to the pioneering work of Fletcher \& Lun on the Kerr metric \cite{fletcherKerrSpacetimeGeneralized2003} and the following expansion by Houque \&  Virmani to the Kerr--de~Sitter solution \cite{hoqueKerrdeSitterSpacetime2021}, we begin by considering Zero Angular Momentum Null Geodesics (ZANGs)  in the Kerr--Newman spacetime in Boyer--Lindquist coordinates \footnote{To maintain a consistent nomenclature with the existing literature, we resolve to use the same notation adopted by Fletcher \& Lun and Houque \&  Virmani \cite{fletcherKerrSpacetimeGeneralized2003,hoqueKerrdeSitterSpacetime2021}.}. These are 
\begin{align} 
    & \B{\rho}^2 \thin \dv{\B{t}}{\lambda} = \frac{\B{\Sigma}^2}{\B{\Delta}} E \,, 
     \label{Geod1}\\
    & \B{\rho}^4 \thin \qty( \dv{\B{r}}{\lambda})^2 = \B{B}^2 \thin  E^2 \,,\label{Geod2}\\
    & \B{\rho}^4 \thin \qty( \dv{\B{\theta}}{\lambda} )^2 =  \B{\Omega} \,,\label{Geod3}\\
    & \B{\rho}^2 \, \dv{\B{\phi}}{\lambda} = \frac{ a \thin \B{\Xi} } {\B{\Delta}}E \,, \label{Geod4} 
\end{align}
where $\lambda$ is an affine parameter along the ZANGs, $E$ is the constant of motion interpreted as the energy of the photons, and the remaining functions appearing in equations \eqref{Geod1}--\eqref{Geod4} are 
\begin{align}
    & \B{\Sigma}^2(\B{r},\B{\theta}) = \B{A}^4 - a^2 \thin \B{\Delta} \sin^2\B{\theta} \,,\\
    & \B{B}^2(\B{r}) = \B{A}^4 - a^2 \thin \B{X}^2 \thin \B{\Delta} \,, \\
    & \B{\Omega}(\B{r},\B{\theta}) = a^2 \thin E^2 \thin (\B{X}^2 - \sin^2\B{\theta} ) \,.
\end{align}
$\B{X} = \B{X}(\B{r},\B{\theta})$ is related to Carter's separation constant, $K$, by $K =a^2 \thin E^2 \thin \B{X}^2$ \cite{CarterGlobalStructure1968}. Hence, it also results as a constant of geodesic motion
\begin{equation}
    \frac{d}{d\lambda} \B{X}(\B{r},\B{\theta}) = 0 \,. \label{MyConstant}
\end{equation}
When $Q = 0$, equations \eqref{Geod1}--\eqref{Geod4} reduce to equations (12)--(15) of \cite{fletcherKerrSpacetimeGeneralized2003}, as it should be expected. Moreover, we point out that as for the pure Kerr case solution, the r.h.s. of \eqref{Geod4} is a function of $\B{r}$ alone. We can now proceed to write the Kerr--Newman metric in GBS coordinates. We start with the coordinate transformation
\begin{align}
    & \B{t} = \T{u} + J(\T{r},\T{\theta)} \, , \label{GBS_1}\\
    & \B{r} = \T{r} \, , \label{GBS_2}\\
    & \B{\theta} = \B{\theta}(\T{r}, \T{\theta}) \, , \label{GBS_3}\\
    & \B{\phi} = \T{\phi} + L(\T{r},\T{\theta}) \, , \label{GBS_4}
\end{align}
where the functions $J(\T{r},\T{\theta)}$, $\B{\theta}(\T{r}, \T{\theta})$ and $L(\T{r},\T{\theta})$ are arbitrarily defined, at this stage. The coordinate transform is chosen in this manner as to preserve the simple form of the killing vector fields in the new coordinate system
\begin{align}
    &\partial_\B{t} = \partial_\T{u} \, , \label{Killing_1}\\
    &\partial_\B{\phi} = \partial_\T{\phi} \,. \label{Killing_2}
\end{align}
We further impose that the integral curves of the ZANGs, in the new coordinates, are lines of constant $\{\T{v},\T{\theta},\T{\phi}\}$, i.e.
\begin{align}
    & \dv{\T{v}}{\lambda} = 0 \, , \label{Int_1}\\
    & \dv{\T{\theta}}{\lambda} = 0 \, , \label{Int_2}\\ 
    & \dv{\T{\phi}}{\lambda} = 0 \, . \label{Int_3}
\end{align}
Applying the coordinate transformation \eqref{GBS_1}--\eqref{GBS_4} with conditions \eqref{Int_1}--\eqref{Int_3} to \eqref{Geod1}--\eqref{Geod4} gives
\begin{align}
    & \pdv{J}{\T{r}}  = \frac{\B{\Sigma}^2}{\B{B}\B{\Delta}} \, , \label{J}\\
    & \pdv{L}{\T{r}}  = \frac{a \thin \B{\Xi}}{\B{\Delta} \thin \B{B}} \, , \label{L}\\ 
    & \qty( \pdv{\B{\theta}}{\T{r}})^2  = \frac{\B{\Omega}}{\B{B}^2 E^2 } \, , \label{dThetadr}
\end{align}
with
\begin{equation}
    \dv{\T{r}}{\lambda} = \dv{\B{r}}{\lambda} = \frac{\B{B} \thin E}{\B{\rho}^2} \,. \label{drdlambda}
\end{equation}
The choice of the positive root for $\T{B}$, combined with \eqref{GBS_1} and \eqref{J}, indicates that we are using a retarded time coordinate and that the ZANGs are outgoing rather than ingoing null geodesics. Furthermore, from \eqref{MyConstant} and \eqref{Int_2} we deduce,
\begin{equation}
    \B{X} = \T{X}(\T{\theta}) \label{XTheta} \, .
\end{equation}
Since we have picked $\B{r} = \T{r}$ we can take the square root of \eqref{dThetadr} and integrate to obtain 
\begin{equation}
    \int^{\B{\theta}} \frac{\dd{\theta'}}{\sqrt{X^2 - \sin^2 \theta'}} = \pm \int^\T{r} \frac{a \dd{r'}}{\T{B}(r')} =: \pm \thin \alpha_X \thin,\label{def_Alpha}
\end{equation}
where
\begin{equation}
    \alpha_X (\T{r}) = \bigintss^{\thin \T{r}} \frac{a \dd{r}'}{ \sqrt{(r^{'2} + a^2)^2 - a^2 \thin X^2 \thin (r^{'2} +a^2 -2 \thin M r' + Q^2 )}}\thin, \label{alpha}
\end{equation}
and
\begin{equation}
    \dv{\alpha_X}{\T{r}} = \frac{a}{B(\T{r})} \thin . \label{der_alpha}
\end{equation}
Here, when $a$ is positive, $\alpha_X$, is chosen to be a negative, monotonically increasing function. To integrate \eqref{def_Alpha} we notice that the l.h.s. is the Legendre incomplete integral of the first kind and hence defines the Jacobi elliptic sine (sn) function. Thus, we have
\begin{equation} \label{eq:sintheta}
    \sin{\B{\theta}} = 
        \begin{cases}
        \rm{sn} \thin \qty(\pm \alpha_X \thin X + H(\T{\theta}), \frac{1}{X^2}) &  X^2 >1 \\
        \tanh(\pm \alpha_X + \rm{H}(\T{\theta})) &   X^2 = 1 \\
        X\thin \rm{sn} (\pm \alpha_X + H(\T{\theta}), \thin X^2 ) &   \sin^2 \B{\theta} \leq X^2 <1
        \end{cases} \, ,
\end{equation}
where $\rm{H}(\T{\theta})$ is an arbitrary function of $\T{\theta}$. We now require $\B{\theta} \rightarrow \T{\theta}$ for $\T{r} \rightarrow \infty$, that is, the two angular coordinates must match at large distances. Therefore, we obtain
\begin{equation}\label{eq:Harb}
    \rm{H} = 
    \begin{cases}
        \rm{sn}^{-1} (\sin{\T{\theta}}, \frac{1}{X^2} ) & X^2 > 1 \\
        \tanh^{-1} (\sin{\T{\theta}}) & X^2 = 1 \\
        \rm{sn}^{-1} \qty(\frac{\sin{\T{\theta}}}{X}, X^2) & \sin^2 \B{\theta} \leq X^2 < 1
    \end{cases} \, .
\end{equation}
Finally, by requiring a fixed equatorial plane under the transformation of coordinates
-- $\B{\theta}= \pm \thin \pi/2 \longleftrightarrow \T{\theta} = \pm  \thin \pi/2$ -- the case $X^2 = 1$ is selected\footnote{Henceforth, the subscript $X$ is dropped from $\alpha_X$.}. This corresponds to choosing the simplest possible class of ZANGs with non-zero energy. Indeed, it forces both Carter's constant ($\mathcal{Q} = K - a^2 \thin E^2$ ) and the total angular momentum about the axis of symmetry to be zero. 
\\ \\
Here, we must also stress an interesting difference between the coordinate systems built following this procedure for the Kerr, Kerr--de~Sitter and Kerr--Newman spacetimes. For the latter, due to the presence of the charge term in the denominator of \eqref{alpha}, the coordinate system is not well-defined over the whole spacetime. Indeed, by analysing \eqref{alpha} for $X = 1$, we see that the coordinate chart develops a singularity at the real positive root of 
\begin{equation}
    P(r;M;a;Q) = (r^{2} + a^2)^2 - a^2 (r^{2} +a^2 -2 \thin M r + Q^2 ) \,.
\end{equation}
Therefore, in the Kerr--Newman case, the coordinate system will not be global, unlike in Kerr and Kerr--de~Sitter. However, the coordinate singularity appears only below the outer horizon of the charged BH. Thus, the coordinate chart built using ZANGs can still be used in studying the asymptotic structure of the spacetime. Then, by using equations \eqref{eq:sintheta} and \eqref{eq:Harb} we obtain
\begin{equation}
    \tanh^{-1} (\sin{\B{\theta}}) = \tanh^{-1}(\sin{\T{\theta}}) \pm \alpha \, , \label{killme}
\end{equation}
with 
\begin{equation}
    \alpha(\T{r}) = - \bigintss_{\T{r}}^\infty \frac{a \dd{r}'}{\sqrt{r^{'4} + a^2(r^{'2} +  r' 2 \thin M -  Q^2)}} \,.
\end{equation}
From \eqref{killme}, and choosing the plus side in front of $\alpha$, we directly deduce
\begin{align}
    & \sin{\B{\theta}} = \frac{D}{C}\, , \label{seno}\\ 
    & \cos{\B{\theta}} = \frac{\cos{\T{\theta}}}{C \thin \cosh{\alpha} }\, , \label{coseno}
\end{align}
where
\begin{align}
    & C = 1 + \tanh{\alpha} \thin \sin{\T{\theta}}\, , \label{C} \\
    & D = \tanh{\alpha} + \sin{\T{\theta}} \, . \label{D}
\end{align}
From \eqref{seno}, \eqref{coseno}, \eqref{C} and \eqref{D} we obtain
\begin{align}
    & \frac{\partial \B{\theta}}{\partial \T{r}} =  \frac{\cos \T{\theta}}{C \cosh \alpha}\frac{\dd \alpha}{\dd \T{r}} = \frac{\cos \T{\theta}}{C \cosh \alpha}\frac{a}{B(\T{r})} \,,  \label{agg1}\\ 
    & \frac{\partial \B{\theta}}{\partial \T{\theta}} =  \frac{1}{C \cosh \alpha} \,. \label{agg2}
\end{align}
Therefore, we have
\begin{align}
    & \dd{\B{t}} = \dd{\T{u}} + \frac{\T{\Sigma}^2}{\T{B} \thin \T{\Delta}} \dd{\T{r}} + g(\T{r},\T{\theta}) \dd{\T{\theta}}\, , \label{dt}\\
    & \dd{\B{\phi}} = \dd{\T{\phi}} + \frac{a \thin \T{\Xi}}{\T{B} \thin \T{\Delta}} \dd{\T{r}} + h(\T{r},\T{\theta}) \dd{\T{\theta}}\, , \label{dphi}\\
    &  \dd{\B{\theta}} =\frac{\cos \T{\theta}}{C \cosh \alpha}\frac{a}{B} \dd{\T{r}} + \frac{1}{C \cosh \alpha} \dd{\T{\theta}} \, , \label{dtheta}
\end{align}
where $g(\T{r},\T{\theta}) = \partial J(\T{r},\T{\theta}) / \partial \T{\theta}$ and $h(\T{r},\T{\theta}) = \partial L(\T{r},\T{\theta}) / \partial \T{\theta}$. To complete the coordinate transformation, we need to establish the function form of $g(\T{r},\T{\theta})$ and $h(\T{r},\T{\theta})$. From the condition
\begin{equation}
    g{_\T{r}}{_\T{\theta}} = 0 \, , \label{condimento}
\end{equation}
we deduce the form of $g(\T{r},\T{\theta})$ as 
\begin{equation}
    g(\T{r},\T{\theta}) = \frac{a \thin \cos{\T{\theta}}}{C^2 \thin \cosh^2 \alpha} \,, \label{salsa}
\end{equation}
whilst from the integrability condition
\begin{equation}
    \partial_\T{\theta} \thin \partial_{\T{r}} \thin L(\T{r},\T{\theta}) = \partial_\T{r} \thin \partial_\T{\theta} \thin L(\T{r},\T{\theta}) \,,
\end{equation}
we get
\begin{equation}
    h(\T{r},\T{\theta}) = h(\T{\theta}).
\end{equation}
Without losing any generality we are then free to choose
\begin{equation}
    h(\T{\theta}) = 0. \label{choiceofH}
\end{equation}
Therefore, \eqref{J}, \eqref{L}, \eqref{agg1}, \eqref{agg2}, \eqref{salsa} and \eqref{choiceofH} completely define the correct coordinate transform -- \eqref{GBS_1}-\eqref{GBS_4} -- to cast 
the line element \eqref{eq:Kerr--Newman_Boyer--Lindquist} into the GBS form
\begin{align*}
    \dd s^2 = & - \qty( 1- \frac{\T{\Xi}}{\T{\rho}^2}) \dd\T{u}^2 -2\frac{\T{\rho}^2}{\T{B}}\dd\T{u}\dd\T{r}-2\qty( 1- \frac{\T{\Xi}}{\T{\rho}^2} ) \thin  \frac{a \thin \cos\T{\theta} }{C^2 \thin \cosh^2 \alpha}\dd\T{u}\dd\T{\theta} -2\frac{a \thin \T{\Xi}}{\T{\rho}^2} \qty( \frac{D}{C} )^2\dd\T{u}\dd\T{\phi} \\
    &  + \left[\frac{\T{\rho}^2}{C^2 \thin \cosh^2 \alpha}\right. \left. - \qty( 1-\frac{\T{\Xi}}{\T{\rho}^2} )\thin \frac{a^2 \thin \cos^2 \T{\theta}}{C^4 \thin \cosh^4 \alpha}\right]\dd\T{\theta}^2 - 2\frac{a^2 \thin \cos{\T{\theta}} }{C^2 \thin \cosh^2 \alpha } \qty( \frac{D}{C} )^2 \frac{\T{\Delta}}{\T{\rho}^2}\dd\T{\theta}\dd\T{\phi}  \\ 
    &+ \qty( \frac{D}{C} )^2 \thin \frac{\T{\Sigma}^2}{\T{\rho}^2} \dd\T{\phi}^2 \,. \numberthis \label{eq:BSform_metric}
\end{align*}
Finally, to put the line element \eqref{eq:BSform_metric} into the BS gauge we apply the following coordinate transformation \cite{Gaur2024}
\begin{align}\label{redefinedr}
    &\T{u} = u \, ,\\
    &\T{\theta} = \theta \, ,\\ 
    &\T{\phi} = \phi \, ,\\
    &\T{r} = r + \frac{a}{2}\frac{\cos2 \theta}{\sin\theta} + \frac{a^2}{8}\,\Bigg(4\cos 2 \theta +\frac{1}{\sin^2\theta}\Bigg) \frac{1}{r}\,. 
\end{align}
At the expansion order of interest in $r$, we find the metric components to be\footnote{The calculations put forward in this paper have been checked with the Mathemathica codes described in the Appendix.}

\begin{align}
    & g_{uu} = -1+\frac{2 M}{r}- \frac{ a M \csc\theta \cos 2 \theta + Q^2}{r^2} + \mathcal{O}\left(r^{-3}\right) \, ,\\
    & g_{ur} = -1 +\frac{a^2 \csc^2\theta }{8 \thin r^2}+\frac{a^2 \left(2\thin M + a\cos 4\theta\csc\theta\right)}{2 \thin r^3} + \mathcal{O}\left(r^{-4}\right) \, ,\\
    & g_{u \phi} = -\frac{2\thin a \thin M \sin ^2 \theta }{r}  + \frac{a \sin \theta  \left(3 a \thin M \cos 2 \theta +2 a M+Q^2 \sin \theta \right)}{r^2} + \mathcal{O}\left(r^{-3}\right) \, ,\\
    & g_{u\theta} = \frac{1}{2} a \cot \theta \csc \theta +\frac{a \cos \theta  \left(a \csc ^3 \theta +8 M\right)}{4 \thin r} - \frac{ a \cot \theta  \csc \theta \left(4 \thin a^2+Q^2\right) \cos 2 \theta }{2 \thin  r^2} \\
    &\quad ~~~~  -\frac{a \cot \theta \csc \theta \left(3 \thin a^2 \cos 4 \theta +2 \left(2 \thin a^2-7 a M \sin \theta +3 \thin a M \sin 3 \theta +Q^2\right)\right)}{4 \thin r^2} + \mathcal{O}\left( r^{-3} \right) \nonumber \,,\\
    & g_{\theta \theta} = r^2+a \thin r \csc \theta +\frac{1}{2} a^2 \csc ^2 \theta +\frac{a^2 \left( a \csc \theta  \cos 4 \theta + 8 M \cos ^2 \theta \right)}{4 \thin r} + \mathcal{O}\left(r^{-2}\right) \, ,\\
    & g_{\theta \phi} = -\frac{2 \thin a^2 M \sin ^2 \theta  \cos \theta}{r}  + \frac{a^2 \sin \theta  \cos \theta  \left(5 \thin a \thin M \cos 2 \theta +Q^2 \sin \theta \right)}{r^2} + \mathcal{O}\left(r^{-3}\right) \, ,\\
    & g_{\phi \phi} = r^2 \thin \sin ^2\theta  -a \thin r \sin \theta  + \frac{a^2}{2} +
     \frac{a^3 \sin \theta \cot ^2 \theta  (\cos 4 \theta -2 \cos 2 \theta )  }{4\thin r}  \nonumber \\
    &  \quad ~~~~ + \frac{a^2 \sin \theta \left(4 \thin a \sin ^4 \theta -5 \thin a \sin ^2 \theta +a+2\thin M \sin ^3 \theta \right)}{r} + \mathcal{O}\left(r^{-2}\right)  \, .
\end{align}



Furthermore, we can compute the electromagnetic four-potential in the selected gauge. We start by considering the four-potential in Boyer--Lindquist coordinates
\begin{equation}
    A_\mu \thin \dd \B{x}^\mu = \frac{\Bar{r}Q}{\Bar{\rho}^2} \dd \B{t} +  a \frac{\Bar{r}Q}{\Bar{\rho}^2}\sin ^2 \Bar{\theta} \thin \dd \B{\phi} \, .
\end{equation}
We then find for the GBS coordinates
\begin{equation}
    A_\mu \thin \dd \T{x}^\mu = \frac{\T{r} \thin Q}{{\T{\rho}}^2} \dd \T{u} + \frac{\T{r} \thin Q}{\T{B} \T{\Delta} }(\T{r}^2+a^2) \thin \dd \T{r} + \frac{\T{r} \thin Q}{{\T{\rho}}^2} \thin \frac{a \cos\T{\theta}}{C^2 \cosh^2 \alpha } \thin \dd \T{\theta} + a \frac{\T{r} \thin Q}{\T{\rho}^2}\left(\frac{D}{C}\right)^2 \thin \dd \T{\phi} \, ,
\end{equation}
where $\T{\rho} = \T{r}^2 + a^2 - a^2\left(D/C \right)^2$. Given that $A_{\T{r}}$ is solely a function of $\T{r}$, it can be set to zero via a classical $U(1)$ gauge transformation. Moving into the BS gauge we find
\begin{align}
    A_\mu \thin \dd x^\mu =& \left(\frac{Q}{r} -\frac{a \thin Q \cos{2 \theta}  \csc \theta }{2 r^2}\right) \dd u + \left(\frac{a \thin Q \cos{\theta }}{r} + \frac{a^2 Q \csc \theta (\cos{\theta}-3 \cos{3 \theta})  }{4 r^2}\right) \thin \dd \theta \nonumber \\ &+ \left(\frac{a \thin Q \sin ^2 \theta }{r} -\frac{a^2 Q \sin{\theta} (3 \cos{2 \theta}+2)}{2 r^2}\right) \thin \dd {\phi} \, + \mathcal{O}(r^{-3})\,.
\end{align}
Thus, we can now move to the evaluation of the asymptotic structure of the spacetime.

\section{The Memory Effect at Null Infinity}\label{sec:3}

The gravitational memory effect as seen by an observer at null infinity has been shown to be equivalent to a BMS supertranslation. Following the investigation of the Kerr memory effect at null infinity \cite{Gaur2024}, we now focus on the Kerr--Newman memory effect. The supertranslated metric functions are calculated via 
\begin{equation}
    \delta g_{\mu\nu} = \mathcal{L}_{\mathbf{\xi}}g_{\mu\nu},
\end{equation}
where $\xi^{\alpha}$ is the asymptotic Killing vector, \eqref{Asymptotic_Killing_vec_truncated}. We find\footnote{By comparison to the Kerr spacetime \cite{Gaur2024}, there are extra orders in the expansion. This is so one can observe the change in the charge.}
\allowdisplaybreaks
\begin{align}\label{supertranslated_kerr_functions}
        &\delta g_{uu} = \frac{1}{r^3}\Bigg\{-Mr + Q^2+ \frac{aM(1-2\sin^2\theta)}{\sin\theta}\Bigg\}D^2f  \nonumber\\
        &\hspace{5.8em}+\frac{aM}{r^3}\Bigg\{(-2 + \cos{2 \theta}) \cot{\theta} \csc{\theta}\Bigg\}\partial_{\theta}f + \laund{r^{-4}}\,,\\
        &\delta g_{ur} =\frac{1}{r^2}\Bigg\{\frac{\thin a\cos\theta}{2\thin\sin^2\theta}\Bigg\}\thin D^2f\,,\\
        &\delta g_{u\thin\theta} =-\Bigg\{\partial_{\theta}f +\frac{1}{2}\partial_{\theta}D^2f\Bigg\} +\frac{1}{r}\Bigg\{2M\partial_{\theta}f - \partial_{\theta} \thin\Bigg(\frac{a}{2}\frac{\cos{\theta}}{\sin^2\theta}\partial_{\theta}f\Bigg)\Bigg\} - \frac{1}{r^2}\Bigg\{Q^2\partial_{\theta}f\Bigg\} \nonumber\\
        &\hspace{11.5em}+\frac{1}{r^2}\Bigg\{aM\csc\theta\cos2\theta\Bigg\}\partial_{\theta}f + \mathcal{O}(r^{-3})\,;\\
        &\delta g_{u\thin\phi} =  -\Bigg\{\partial_{\phi}f + \frac{1}{2}\partial_{\phi}D^2f\Bigg\}+ \frac{1}{r}\Bigg\{2M\partial_{\phi}f - \frac{a}{2}\frac{\cos{\theta}}{\sin^2\theta}\partial_{\phi}\thin\partial_{\theta}f\Bigg\}- \frac{1}{r^2}\Bigg\{Q^2\partial_{\phi}f\Bigg\} \nonumber\\
        &\hspace{11.8em}+\frac{1}{r^2}\Bigg\{aM\csc\theta\cos2\theta\Bigg\}\partial_{\phi}f +\mathcal{O}(r^{-3})\,;\\
        &\delta g_{\theta\theta} =  \Bigg\{2\partial_{\theta}{^2}f - D^2f\Bigg\}\thin r - \frac{a}{\sin\theta}\Bigg\{+\frac{1}{2}D^2f + 2\thin\frac{\cos\theta}{\sin\theta}\partial_{\theta}f - 2\thin\partial_{\theta}{^2}f\Bigg\} + \laund{r^-1}\,;\\
        &\delta g_{\phi\phi}= \Bigg\{ 2\partial_{\phi}{^2}f +2\thin\sin\theta\cos\theta\,\partial_{\phi}f- \sin^2\theta D^2f\Bigg\}\thin r \nonumber\\
        &\hspace{11.9em}-\frac{a}{\sin\theta}\Bigg\{\frac{1}{2}\sin^2\theta D^2f + \cos\theta\thin\partial_{\phi}f + 2\thin\partial_{\phi}{^2}f\Bigg\} +\laund{r^{-1}}\,.
\end{align}
Additionally, the supertranslated gauge field components at null infinity are:
\begin{align}\label{STgaugefield_null_inf}
    &\delta A_{u} =-\frac{1}{2}\frac{Q}{r^2}D^2f -\frac{1}{r^2}\csc{\theta} \Big(\cos{2\theta} \cot{\theta} + 2 \sin{2\theta} \Big) \partial_{\theta}f + \mathcal{O}(r^{-3})\,;\\
    &\delta A_{B} = \frac{1}{2}D^2f\partial_{r}A_{B} - \partial_{\theta}f\partial_{\theta}A_{B} + A_{C}\partial_{B}D^{C}f\\
    &\hspace{5em}+\Bigg(\frac{Q}{r} -\frac{a \thin Q \cos{2 \theta}  \csc \theta }{2 r^2}\Bigg)\partial_{\theta}f  + \mathcal{O}(r^{-3})\nonumber \,.
\end{align}
As can be seen, the supertranslated gauge field has components which match the leading order parts of the original gauge field. Therefore, when an observer at null infinity measures the Maxwell field through $F_{\mu\nu}$, they will observe a difference in a bald Kerr--Newman spacetime and hairy Kerr--Newman spacetime. 

Comparing the new supertranslated metric components with \eqref{generalBMsexpansion} we find $C_{AB}C^{AB}$, $C_{AB}$, $N_{A}$, and $m_{\text{bondi}}$, after the impact of the gravitational wave\footnote{There have been developments in the BMS group where authors have started investigating higher order terms in the expansion. For instance, in \cite{lambda_Compere,kdsinbondi} there are higher order terms, such as `$E_{AB}$' and `$F_{AB}$' which appear in the $g_{AB}$ expansion. However, these modifications are made for the inclusion of a cosmological constant. The relevance of these terms in our analysis and the effect these may have on charges that we observe at null infinity remains unclear and is perhaps an avenue for further research. Furthermore, the incorporation of these terms would likely also require tweaking of the transformation \eqref{redefinedr}, similar to what is done in \cite{kdsinbondi}.}.
\allowdisplaybreaks
\begin{align}\label{supertranslated_various_comps}
       & C_{AB}\thin C^{AB} = \frac{2\thin a^2}{\sin^2\theta} + \frac{16\thin a\cos\theta}{\sin^2\theta}D^2f \,,\\
       & C_{AB}\thin \dd x^{A}\thin \dd x^{B} = \Bigg(\frac{a}{\sin\theta} + 2\partial_{\theta}{^2}f - D^2f\Bigg)\,\dd \theta^2 \nonumber\\
       & \hspace{5.7em}-\Bigg( a\sin\theta - 2\partial_{\phi}{^2}f -2\thin\frac{\cos\theta}{\sin\theta}\partial_{\phi}f- \sin^2\theta D^2f\Bigg)\, \dd \phi^2 \,,\\
       & N_{\theta} = 3M\{\thin a\cos\theta + \partial_{\theta}f\thin \} + \frac{3}{2}\thin a\thin\partial_{\theta}\Bigg\{\frac{\cos\theta}{\sin^2\theta}\Big[D^2f - \frac{1}{2}\partial_{\theta}f\Big] \Bigg\} \,,\\
       &N_{\phi} = 3M\{\thin- a\sin^2\theta + \partial_{\phi}f\thin \} + \frac{3}{2}\thin a\thin\partial_{\phi}\Bigg\{\frac{\cos\theta}{\sin^2\theta}\Big[D^2f - \frac{1}{2}\partial_{\theta}f\Big] \Bigg\}\,,\\
       & m_{\text{bondi}} = M \,.
\end{align}

We are now in a position to discuss the supertranslation and superrotation charges that are implanted on the BH horizon, as seen by an observer at null infinity. As expected \cite{HPS,Gaur2024}, the scattering of a gravitational wave by the BH will not excite supertranslation charge. However, this process, equivalent to a supertranslation at null infinity, will modify the superrotation charge. The superrotation charge that is measured at future null infinity is given by
\begin{equation}
    Q_Y = \frac{1}{8\pi}\int_{\mathcal{I}^+} \dd^2\Theta \,\sqrt{\gamma}\, Y^A N_A\,.
\end{equation}
Using \eqref{supertranslated_various_comps} we get 
\begin{equation}\label{Superrotation_theta_charge}
    \begin{aligned}
    Q_{Y=Y^{\theta}} = \, &\frac{1}{8\pi}\int_{\mathcal{I}^+}  \sqrt{\gamma}\,\dd^2\Theta\,Y^{\theta}\,3M a\cos\theta\,+\\
    & \frac{1}{8\pi}\int_{\mathcal{I}^+}\sqrt{\gamma}\,\dd^2\Theta \,Y^{\theta}\Bigg[\partial_{\theta}f + \frac{3}{2}\thin a\thin\partial_{\theta}\Bigg\{\frac{\cos\theta}{\sin^2\theta}\Big[D^2f - \frac{1}{2}\partial_{\theta}f\Big] \Bigg\}\Bigg]\,,
    \end{aligned}
\end{equation}
and 
\begin{equation}\label{superrotation_phi_charge}
    \begin{aligned}
        Q_{Y=Y^{\phi}} =\,  &\frac{1}{8\pi}\int_{\mathcal{I}^+}-\sqrt{\gamma}\,\dd^2\Theta\,Y^{\phi}\,3M a\sin^2\theta\, + \\
         &  \frac{1}{8\pi}\int_{\mathcal{I}^+}\sqrt{\gamma}\,\dd^2\Theta \,Y^{\phi}\Bigg[\partial_{\phi}f + \frac{3}{2}\thin a\thin\partial_{\phi}\Bigg\{\frac{\cos\theta}{\sin^2\theta}\Big[D^2f - \frac{1}{2}\partial_{\phi}f\Big] \Bigg\}\Bigg]\,.
    \end{aligned}
\end{equation}

The first terms in \eqref{Superrotation_theta_charge} and \eqref{superrotation_phi_charge} correspond to the bald Kerr--Newman BH superrotation charges and can easily be recovered if the supertranslation function $f$ vanishes. Furthermore, when $a=0$ we recover the superrotation charges of the hairy Schwarzschild BH \cite{HPSrot,cit:bhm,KKSofthair}. Moreover, as shown by Barnich and Troessaert in \cite{Barnich_troes}, when $Y^{\phi}$ is the Killing vector, $\partial_{\phi}$, \eqref{superrotation_phi_charge} corresponds to conservation of angular momentum for both the bald Schwarzschild and Kerr BH. For the hairy  Kerr--Newman BH, one may see that the zero-mode superrotation charge (when $f=0$ and $Y^{\phi} = 1$) given by \eqref{superrotation_phi_charge}, does not change and will still correspond to the conservation of angular momentum.

We note that the calculated charges are no different from those obtained for the Kerr BH \cite{Gaur2024}. Therefore, within the current framework, the expected memory effect at null infinity in these two spacetimes is indistinguishable at the level of the metric. This follows from the electric charge, $Q$, appearing only at a higher order than $r^{-1}$ in the expansion of the metric in the BS gauge. In our opinion, this result represents a clear drawback of the current, first-order framework. A higher-order approach is needed to distinguish fundamentally different spacetimes, such as the Kerr and Kerr--Newman solutions, and should therefore be pursued as an important milestone for the field \cite{Compere:2019bua}.

Nonetheless, we point out that the presence of an electromagnetic field in the Kerr--Newman spacetimes gives a novel method to measure the scattering of a gravitational wave from the BH, via the change in the field. In particular, if such a change were to be detected and agree with our calculations, it could be considered as an indirect test for the presence of supertransformation charges. However, such a measurement clearly presents observational challenges not easily met.

\section{Near Horizon Physics: Extremal Kerr--Newman}\label{sec:4}

We now shift our attention to the NH form of the Kerr--Newman spacetime. In particular, contrary to the null infinity analysis, we show that the two spacetimes differ in their response to the scattering of a gravitational shockwave. Indeed, in the Kerr--Newman case, the gravitational wave excites supertransformation charges and implants soft, electric hair on the horizon, due to its interaction with the electromagnetic four-potential. To determine the charges that are implanted on the horizon, we must first find the NH metric components, and secondly, derive an expression for the electromagnetic gauge field in the NH limit. Chrusciel \cite{longgrnotes} shows that the general form of a NH metric is given by
\begin{equation}\label{eq:general_near_horizon}
    \dd s^2 = -2R\thin \kappa \thin \dd v^2 + 2\thin \dd v\thin \dd R +2R \thin \theta_A \dd v \thin \dd x^A + \Omega_{AB} \thin \dd x^A \dd x^B + ...\,,
\end{equation}
where $v$ is the advanced time, $x^A$ are angular coordinates, $\theta_A$, $\Omega_{AB}$ $\equiv\Omega_{a}\,\gamma^{\thin a}{_{AB}}$\footnote{Note the use of the \textit{internal} index, $a$, here. This is required in the case of the NH Kerr--Newman metric as we can not use only one scaling factor for $\Omega_{\Theta\Theta}$ and $\Omega_{\Phi\Phi}$.} are in principle arbitrary metric functions of $v$ and $x^A$, and $\kappa$ is the surface gravity. Note, that when dealing with an extremal horizon --- as is the case in this paper --- the surface gravity vanishes, i.e., $\kappa = 0 $. In the coordinate used in \eqref{eq:general_near_horizon}, the horizon is now located at $R = 0$ and the ellipsis are to denote terms that are $\mathcal{O}(R^2)$. Furthermore, we have the constraints
\begin{equation}\label{eq:near_horizon_constraint1}
    \begin{aligned}
    g_{RR}=0\,,\quad g_{v R}=1\,,\quad g_{AR}=0\,.
    \end{aligned}
\end{equation}
Additionally, in analogy to \cite{donnaysuperathorizon,black_hole_memory_og}, we use the boundary conditions
\begin{equation}\label{eq:near_horizon_constraint2}
        g_{vv} = -2\kappa\thin  R + \mathcal{O}(R^2)\,, \quad g_{vA} = \theta_A R +\mathcal{O}(R ^2)\,, \quad g_{AB} = \Omega_{AB} + \mathcal{O}(R)\,.
\end{equation}
We can then find a set of asymptotic Killing vectors that preserve \eqref{eq:near_horizon_constraint1} and \eqref{eq:near_horizon_constraint2}, generating an algebra consisting of \textit{both} supertranslations and superrotations. The resulting Killing vectors are
\begin{equation}\label{eq:near_horizon_killing_vect}
    \xi^\mu \partial_{\mu} = f\partial_v + \Big( Y^A - \partial_Bf \int^\rho d\rho' g^{AB} \Big)\partial_A + \Big(Z(v,x^A) -\rho \partial_vf + \partial_Af \int^\rho d\rho ' g^{AB}g_{vB} \Big) \partial_{\rho} \,.
\end{equation}
In contrast with \cite{donnaysuperathorizon,black_hole_memory_og}, we find a vector, $Y^A$, that is a ``constant'' of integration which represents the horizon superrotations\footnote{It is important to point out that one does not need a gravitational shockwave here to have ``a superrotation/supertranslation charge''. These aspects exist as a property of the asymptotic structure of the NH metric.}. Then, using the NH asymptotic Killing vector, we compute the general supertranslated metric functions; $\kappa$, $\theta_A$, and $\Omega_{AB}$ subject to \eqref{eq:near_horizon_constraint2} \footnote{We correct a small mistake here that is present in ref \cite{donnaysuperathorizon}. This third equation now correctly states that the Lie derivative of $\Omega_{AB}$ is along $Y^{\alpha}$ and not $\xi^{\alpha}$.}.
\begin{align}
     &\delta_{\xi} \kappa = \mathcal{L}_{\xi} \kappa = 0\,, \label{eq:supertranslated_near_horizon_functions1}\\
     &\delta_{\xi} \theta_A = \mathcal{L}_{Y} \theta_A + f\partial_v \theta_A -2\kappa \partial_A f - 2\partial_v \partial_A f +\Omega^{BC} \partial_v \Omega_{AB} D_C f\,, \label{eq:supertranslated_near_horizon_functions2}\\
     &\delta_{\xi} \Omega_{AB} = f\partial_v \Omega_{AB} + \mathcal{L}_{Y} \Omega_{AB}\,. \label{eq:supertranslated_near_horizon_functions3}
 \end{align}
To properly study the NH physics of a charged BH, we must also discuss the NH expansion of the gauge field. The Taylor expansion of the $\mathrm{U}(1)$ electromagnetic gauge field near $R = 0$ is given by \cite{black_hole_memory_og}
\begin{align}\label{guage_field_expansion}
      &A_v = A^{(0)}_v +R A^{(1)}_v (v,x^A) + \mathcal{O}(R^2),\\ 
      &A_B = A^{(0)}_B (x^A) + R A^{(1)}_B (v,x^A) +\mathcal{O}(R^2)\,,\\
      &A_{R} = 0.
\end{align}
Here $A^{(0)}_v$ is the Coulomb potential. In particular, we find that the supertranslated gauge field components take the form:
\begin{align}\label{gauge_field_diff}
     &\delta_{\xi} A_v = 0,\\
     &\delta_{\xi} A_B = Y^C\partial_C A^{(0)}_B +A^{(0)}_C \partial_B Y^C +\partial_B U.\label{gauge_field_diff_2}
\end{align}
Where $U$ is an arbitrary function of angular coordinates and is referred to as the \textit{electromagnetic charge generator}, just as $f$ is referred to as the generator of supertranslations.

We now discuss the NH extremal Kerr-Newman spacetime and provide the supertranslated metric functions. This will allow us to examine the effect of a gravitational shockwave -- under the identification of supertranslations with the scattering of such waves by the BH -- on the extremal horizon as seen by an observer near the horizon. This leads to a horizon superrotation that is absent at null infinity\footnote{While not discussed in this paper, various parts of the literature make explicit the fact that supertranslations turn on superroation charge and superrotations turn on supertranslation charges. We stated that there is a change in the \textit{superrotation charges} at null infinity due to the \textit{supertranslation}. However, in the near horizon case, (mathematically due to the boundary conditions) we note that there is also a superrotation that has associated \textit{supertranslation charges} discussed in \autoref{NH_charge_section}.}, similarly to the Schwarzschild and Kaluza--Klein cases discussed in \cite{black_hole_memory_og,KKSofthair} respectively. 
\subsection{Near Horizon Metric and Gauge Four-Potential}

To derive the extremal NH Kerr--Newman metric, we begin by defining \cite{longgrnotes}
\begin{align}
    &\Bar{t} = \epsilon^{-1} \hat{t} \,,\\
    &\Bar{r} = M + \epsilon \hat{r} \,,\\
    &\Bar{\theta} = \hat{\theta} \,,\\
    &\Bar{\phi} = \hat{\phi} +\epsilon^{-1}\frac{a}{r_0^2}\hat{t} \,,
\end{align}
where $r_0^2 = M^2 + a^2 $. After taking the limit $\epsilon \rightarrow 0$, the metric becomes
\begin{align}\label{eq:NH_metric_KN}
    ds^2 = &\left(1 - \frac{a^2}{r_0^2}\sin^2 \hat{\theta}\right)\left[-\frac{\hat{r}^2}{r_0^2}d\hat{t}^2 + \frac{r_0^2}{\hat{r}^2}d\hat{r}^2 + r_0^2 d\hat{\theta}^2 \right]  \nonumber \\
    & + r_0^2\sin^2\hat{\theta}\left(1 - \frac{a^2}{r_0^2}\sin^2 \hat{\theta}\right)^{-1} \left[d\hat{\phi} + \frac{2\,a\,M}{r_0^4} \,r\, d\hat{t}\right]^2 \,.
\end{align}
This metric is clearly singular on the horizon. Hence, we apply the following coordinate transform
\begin{align}\label{second_NH_transformation}
    & \hat{t} = V - \frac{r_0^2}{r} \,,\\
    &\hat{r} = R \,,\\
    &\hat{\theta} = \Theta \,,\\
    &\hat{\phi} = \Phi -\frac{2Ma}{r_0^2}\log \left( \frac{\hat{r}}{r_0}\right) \,,
\end{align}
leading to the line element
\begin{align}\label{near_horizon_KN}
    \d s^2 = &\,\frac{\left(r_0^2 - a^2\sin^2 \Theta\right)}{r_0^2}\left[-\frac{R^2}{r_0^2}\d V^2 -2\d V \d R + r_0^2\d\Theta^2 \right] \nonumber \\
    & + \frac{r_0^4\sin^2\Theta}{r_0^2 - a^2\sin^2 \Theta} \left[\d\Phi + \frac{2aM}{r_0^4}R \d V\right]^2 \,,
\end{align}
which is regular for $R = 0$. 
We may now read off the metric functions in \eqref{eq:general_near_horizon}:
\begin{align}
        &\kappa = 0\thin ;\\
        &\theta_{\Theta} = 0\thin ; \\
        &\theta_{\Phi} = \frac{2\,a\,M \thin \sin^2 \Theta}{r_0^2 - a^2 \thin \sin^2 \Theta} ;\\
        &\Omega_{\Theta\Theta} = r_0^2 - a^2 \thin \sin^2 \Theta\thin ;\\
        &\Omega_{\Phi\Phi} = \frac{r_0^4 \thin \sin^2 \Theta}{r_0^2 - a^2 \thin \sin^2 \Theta}\thin . 
\end{align}



We must now bring the Kerr--Newman gauge potential into the form \eqref{guage_field_expansion}.
Performing the same coordinate transformations as we did for the metric, we first find:
\begin{equation}
    A_\mu \thin \dd \hat{x}^\mu = \frac{Q\left(M^2 - a^2\cos^2\hat{\theta}\right)}{r_0^2\left(M^2 + a^2\cos^2\hat{\theta}\right)}\,\hat{r} \thin \thin \dd \hat{t} +  \frac{Q \thin a \thin M\sin^2\hat{\theta}}{M^2 + a^2\cos^2\hat{\theta}} \thin \dd \hat{\phi} \, ,
\end{equation}
where we have eliminated the constant term $\epsilon^{-1}(MQ/r_0^2)$ in $A_{\hat{t}}$ through a $U(1)$ gauge transformation before taking the limit for small $\epsilon$. With the final coordinate transformation \eqref{second_NH_transformation}, we obtain
\begin{equation}
    A_\mu \thin \dd X^\mu = \frac{Q\left(M^2 - a^2\cos^2\Theta\right)}{r_0^2\left(M^2 + a^2\cos^2\Theta\right)}R \thin  \dd V +  \frac{Q \thin a \thin M\sin^2\Theta}{M^2 + a^2\cos^2\Theta} \thin \dd \Phi \, ,
\end{equation}
where we have renormalised $A_R = -(M^2-a^2)(Q/r_0^2) \thin R^{-1}$ with a further $U(1)$ gauge transform. Note, that the Coulomb potential does not appear here. However, because it is coordinate-independent, this can be added back in at any point without changing the Maxwell field. Moreover, as expected, we will see that the Coulomb potential will not appear in the expressions for surface charges.

\subsection{Near Horizon Supertranslations and Charges}\label{NH_charge_section}

Bringing the asymptotic Killing vector, \eqref{Asymptotic_Killing_vec_truncated} to the NH limit for the extremal case -- i.e., $M^2 = a^2+Q^2$ -- and supertranslating the NH Kerr--Newman spacetime \eqref{near_horizon_KN} we find the following metric components
\begin{align}
    g{_V}{_V} &= -\frac{R^2}{r_0^2} \frac{\left(r_0^2 - a^2 \thin \sin^2 \Theta\right)}{r_0^2} + \frac{r_0^4 \thin \sin^2 \Theta}{r_0^2 - a^2 \thin \sin^2 \Theta} \left( \frac{2 a M R}{r_0^4} \right)^2 \,,\\
    g{_\Theta}{_\Theta} &= \Big\{r_0^2 - a^2 \thin \sin^2 \Theta\Big\}\Big\{1 -\frac{2}{r_+}\partial_{\Theta}^2f\Big\} +\frac{1}{r_+}a^2\sin 2\thin \Theta\thin \partial_{\Theta}f \,,\\
    g{_\Phi}{_\Phi} &= \Big\{\frac{1}{r_+}\frac{r_0^4}{r_0^2 - a^2 \thin \sin^2 \Theta}\Big\}\Big\{{r_+\sin^2{\Theta}}-\frac{r_0^2\sin 2\thin \Theta}{r_0^2-a^2\sin^2\Theta}\partial_{\Theta}f - 2\partial_{\phi}^2f\Big\} \,,\\ 
    g{_V}{_R} &= -\frac{r_0^2 - a^2 \thin \sin^2 \Theta}{r_0^2} \,,\\
    g{_V}{_\Phi} &=  \Bigg\{\frac{1}{r_+}\frac{2\,a\,M }{r_0^2 - a^2 \thin \sin^2 \Theta}\Bigg\}\Bigg\{\sin^2\Theta - \frac{\sin\thin 2\thin \Theta\thin r_0^2}{r_0^2 - a^2 \sin^2 \Theta}\partial_{\Theta}f - \partial^2_{\Phi}f\Bigg\}R \,,\\
    g{_V}{_\Theta} &=\Bigg\{\frac{1}{r_+}\frac{2\,a\,M }{r_0^2 - a^2 \thin \sin^2 \Theta}\Bigg\}\Bigg\{2\cot\Theta \thin \partial_{\Phi}f - \partial_{\Theta}\thin \partial_{\Phi}f\Bigg\}\, R  \,,\\
    g{_R}{_R} &= 0 \,,\\
    g{_R}{_\Theta} &= 0 \,,\\
    g{_R}{_\Phi} &= 0 \,,\\
    g{_\Theta}{_\Phi} &= 0 \,.
\end{align}
We can now compare our results with the supertranslated extremal NH Kerr--Newman spacetime, see \eqref{eq:NH_metric_KN}, and the general NH supertranslated metric functions, see \eqref{eq:supertranslated_near_horizon_functions1}-\eqref{eq:supertranslated_near_horizon_functions3}.
Since $\Omega_{AB}$ does not depend on retarded/advanced time, from \eqref{eq:supertranslated_near_horizon_functions3} we find the corresponding horizon superrotation to be
\begin{equation}\label{near_horizon_superrot}
    Y_A = \frac{1}{M} D_Af\,.
\end{equation}
Surprisingly, this does not differ from the horizon superrotation of a Schwarzschild BH found in \cite{black_hole_memory_og}. Indeed, this is perhaps not expected as the Kerr class of solutions are already rotating. However, this could be intuitively understood by noticing that the correlation between the memory effect and supertranslations --- in both regimes, null infinity, and NH --- is only examined at linear order. In fact, one can show that if $\Omega_{AB}$ does not depend on advanced or retarded time, we will always have a horizon superrotation of this form --- up to a factor which depends on the horizon radius.

The associated charges with the diffeomorphisms generated by asymptotic Killing vectors have associated horizon charges. The derivation of these charges stem from \cite{Barnich_NH_Charges} and are also discussed in \cite{black_hole_memory_og,donnay_nh_sym}. The NH charges take the form: 
\begin{equation}\label{nhchargeeq}
    \mathcal{Q}\thin [X,Y^A,U] = \frac{1}{16\pi}\int \dd \Theta\thin\dd \Phi \sin \Theta\, r_0^2 \Bigg(2X - Y^A\theta_{A} - 4UA_{V}^{(1)} - 4A_{B}^{(0)}Y^{B}A_{V}^{(1)}\Bigg)\,.
\end{equation}
In the extremal case, it is apparent that the surface gravity vanishes and so too does the Hawking temperature \cite{bardeen_four_1973,hawking_particle_1975-1}. This leads to an interesting scenario in which the Hawking temperature is no longer the associated zero-mode for the first charge. In this case, this zero-mode (the supertranslation charge) is associated with the product of Bekenstein-Hawking entropy and the geometric temperature \cite{black_hole_memory_og,Hartman:2008pb}. The second term is analogous to the superrotation charge found at null infinity. The third term is due to the electromagnetic generator and the last term mixes the superrotation vector field with the gauge field. 

Let the associated charges to $X$, $Y^{A}$, and $U$ be $\mathcal{X}$, 
 $\mathcal{Y}^{A}$, and $\mathcal{U}$ respectively. The associated zero-mode (bald) horizon charges are
\begin{align}
        &\mathcal{X} = \mathcal{Q}[1,0,0] = \frac{r_0^2}{2} \,,\\
        &\mathcal{Y}^{\Theta} = \mathcal{Q}[0,Y^{\Theta}=1,Y^{\Phi}=0,0] = 0 \,,\\
        &\mathcal{Y}^{\Phi} = \mathcal{Q}[0,Y^{\Phi}=1,0] = \frac{1}{16\pi}\int \dd \Theta\thin\dd \Phi \sin \Theta\, r_0^2 \Bigg(\theta_{\Phi} - 4A_{\Phi}A_{V}^{(1)}\Bigg)\,,\\
        &\mathcal{U} = \mathcal{Q}[0,0,1] = -\frac{1}{4\pi}\int\dd \Theta \dd\Phi\sin \Theta\, r_0^2 \Big(A_V^{(1)}\Big) \, \nonumber \\
        &\qquad \qquad \qquad = Q\left(1-\frac{2M}{a}\arctan{\left(\frac{a}{M}\right)}\right).
\end{align}
The zero mode of $\mathcal{Y}^{\Phi}$ gives the angular momentum of the BH as measured by the hovering observer. As one may note, there is a contribution to this zero-mode from the gauge field which does not vanish. Therefore, we see a strong interaction between the electromagnetic gauge potential and the angular momentum of the BH, with the former influencing the latter for the chosen observer. As the gauge field vanishes, we retrieve the extremal NH Kerr solution, and the angular momentum depends solely on $\theta_{\Phi}$. The final charge, $\mathcal{U}$, whose zero-mode charge corresponds to the electromagnetic charge generator, gives the total electric charge of the BH as measured in the NH limit. Here, the complementary effect is observed and the angular momentum of the BH effectively shields the intrinsic charge for the NH observers. Remarkably, these unexpected effects of the self-interactions between angular momentum and electric charge are not found via a null infinity analysis. Thus, they further indicate the importance in general relativity of studying the same phenomena using a plurality of observers. Lastly, one may verify that these charges do indeed agree with the extremal Reissner--Nordstr\"om horizon when $a\to 0$ as seen in \cite{black_hole_memory_og}.\\

We may also use \eqref{nhchargeeq} to determine the zero-mode of the NH charges of the supertranslated horizon. To do so, we compute
\begin{align}
    &\theta_{\Theta} = \Bigg\{\frac{1}{M}\frac{2\,a\,M }{r_0^2 - a^2 \thin \sin^2 \Theta}\Bigg\}\Bigg\{2\cot\Theta \thin \partial_{\Phi}f - \partial_{\Theta}\thin \partial_{\Phi}f\Bigg\}\,;\\
    & \theta_{\Phi} = \Bigg\{\frac{1}{M}\frac{2\,a\,M }{r_0^2 - a^2 \thin \sin^2 \Theta}\Bigg\}\Bigg\{\sin^2\Theta - \frac{\sin\thin 2\thin \Theta\thin r_0^2}{r_0^2 - a^2 \sin^2 \Theta}\partial_{\Theta}f - \partial^2_{\Phi}f\Bigg\}\,;\\
    &A_V^{(1)} = \frac{Q}{r_0^2}\Bigg\{\frac{\left(M^2 - a^2\cos^2\Theta\right)}{\left(M^2 + a^2\cos^2\Theta\right)}- \frac{1}{M}\partial_{\Theta}\frac{(M^2 - a^2\cos^2\Theta)}{(M^2 + a^2 \cos^2\Theta)}\partial_{\Theta}f\Bigg\}\,;\\
    &A_B^{(0)} =  \frac{Q \thin a \thin M\sin^2\Theta}{M^2 + a^2\cos^2\Theta} + \delta A_B\,,\\
\end{align}
where $\delta A_{B}$ is given below. Interestingly, we see that once the NH spacetime is supertranslated by the passage of a gravitational wave, then $\theta_{\Theta}$ is no longer zero. However, even though the NH geometry is transformed, the zero-mode horizon charges remain unchanged. This is because we are setting the supertranslation generator, $f$, to zero\footnote{In fact, $f$, can be expanded in Fourier modes which relate it linearly to $X$ in the extremal case. Hence, when setting $X$ to zero, we are also setting $f$ to zero, and the zero-modes now correspond solely to the bald near-horizon geometry.} in all cases.

In the bald and supertranslated metric analysis, we already see an interplay between the electromagnetic field and angular momentum. However, a further interaction between the gravitational and electromagnetic fields can be investigated by determining the change in the electromagnetic field generator due to the memory effect in the NH limit
\begin{equation}
    U = \int \Big( Y^{C}\partial_{C}A^{(0)}_B + A^{(0)}_C\partial_{B}Y^{C} - \delta_{\mathbf{\xi}} A_{B}\Big)\, \dd x^{B}\,.
\end{equation}
Here, $Y^{A}$ is the horizon superrotation,
\begin{equation}
    Y^{A} = \frac{1}{M}D^Af\,,
\end{equation}
and
\begin{equation}\label{deltaAB}
    \delta_{\mathbf{\xi}}A_{B} = -Q\thin a\,\Bigg\{\partial_{\Theta}\Bigg(\frac{\sin^2\Theta}{M^2 + a^2\cos^2\Theta}\Bigg)\partial_{\Theta}f - \Bigg(\frac{1}{M^2 + a^2\cos^2\Theta}\Bigg)\partial_{\Theta}\partial_{\Phi}f\Bigg\} \,.
\end{equation}
This illustrates that the gravitational memory effect due to the passing of a gravitational wave is not only seen as a supertranslation from null infinity, but in the NH limit implants \textit{soft electric hair} on the extremal Kerr--Newman horizon.  

\section{Conclusions}\label{sec:5}

Motivated by the rising relevance of the gravitational memory effect, in this paper we have investigated the connection between the scattering of a gravitational shockwave by the Kerr--Newman black hole, as seen in the near-horizon region and in the far asymptotic region. 

In \cite{HPS} the authors showed that a transient gravitational shockwave modifies the black hole geometry in a way that can be interpreted as a BMS supertranslation at null infinity. Here, we have brought for the first time the Kerr--Newman black hole in the Bondi--Sachs gauge and computed the action of a BMS supertranslation on its asymptotic structure. In particular, we discussed the change in the supertransformation charges due to the supertranslation
hair implanted on the Kerr--Newman black hole by the gravitational wave. We have shown that the supertranslation charge was absent at null infinity, whilst a superrotation charge is instead detectable. Furthermore, the zero-mode of the superrotation charge remains unchanged, as any change in mass cannot be captured by the action of pure BMS supertranslations. 

Following the pioneering work of Donnay \textit{et al} \cite{donnay_nh_sym}, we studied the gravitational memory effect in the near horizon limit of an extremal Kerr--Newman black hole. Surprisingly, we found that the corresponding horizon superrotation matches the one computed for non-rotating black holes. Moreover, we find that no non-trivial supertranslation charge is turned on at the horizon, due to the extremality of the black hole. Finally, we show that the scattering of the gravitational shockwave by the black hole implants soft electric hair on the horizon, via its interaction with the electromagnetic gauge field.

Some questions remain open and require further study. Indeed, we showed that a higher-order formalism is needed to properly capture the full properties of the spacetime when dealing with the memory effect. Consequently, a rigorous definition, and interpretation, of higher-order charges would be required. Furthermore, we have found a series of previously unexplored interactions between the gravitational and electromagnetic fields. To wit, the presence of electric charge invalidates the construction of the Bondi--Sachs coordinates as a global coordinate patch for the spacetime, failing below the horizon. Moreover, we have showed that the electric charge, and angular momentum, inferred for the black hole by a near horizon observer differ from what would be measured in the asymptotic region, on account of the interplay between these two quantities. This unexpected interaction between spin and charge requires further clarification, with a possible avenue of research leading to the study of a similar effect in higher dimensional, charged, rotating black holes.

\acknowledgments
MG and CHH were supported by the University of Canterbury Doctoral Scholarship. RG was supported by a Victoria University of Wellington PhD Doctoral Scholarship. MG and CHH would like to warmly thank Matt Visser for his hospitality during the preparation of this paper. The authors would like to acknowledge Chris Stevens, Matt Visser and David Wiltshire for their numerous comments and insights on this paper.  

\section*{Appendix: Mathematica Codes}

We have submitted three ancillary Mathematica files with this submission to the arXiv. A brief explanation on these files is as follows.
\begin{enumerate}
    \item \textbf{KerrNewmanGeneralisedBondiSachsForm.nb}: this file writes the Kerr--Newmann metric into the generalised Bondi--Sachs form. \\
    \item \textbf{KerrNewmanBondiGaugeComponentsExpansions.nb}: this file computes the Kerr--Newmann metric components expansions at null infinity in the generalised Bondi--Sachs form; transforms the metric into the Bondi--Sachs gauge and computes their asymptotic expansion. \\
    \item \textbf{KerrNewman4PotentialNullInfinity.nb}: this file computes the electromagnetic four-potential in the Bondi--Sachs gauge.
\end{enumerate}

\bibliographystyle{JHEP}
\bibliography{main.bib}

\end{document}